\documentstyle[fullpage,11pt]{article}
\def\la{\langle}
\def\ra{\rangle}

\begin{document}
\newtheorem{theorem}{Theorem}
\newtheorem{statement}{Statement}
\newtheorem{lemma}{Lemma}
\newtheorem{definition}{Definition}
\newtheorem{corollary}{Corollary}

\begin{center}
{\LARGE Introduction to the
functions on compact Riemann surfaces and Theta-functions}
\vskip2.0cm
{\large D.A.Korotkin}\footnote{On leave of absence from Steklov Mathematical
Institute, Fontanka, 25, St.Petersburg 191011, Russia}
\vskip0.7cm
II.Institute for Theoretical Physics,
Luruper Chaussee 149, Hamburg 22671 Germany
\end{center}

{\bf Abstract.}
We collect here some classical facts related to analysis on the Riemann
surfaces. The notes may serve as an introduction to this field;
we suppose that the reader is familiar only with the basic facts
from topology and complex analysis. The  treatment
is organised to give the background for the 
further application to non-linear
differential equations.
The interested reader may refer for more
details the numerous nice books :{\it Springer} {\cite{Spring}},
{\it Hurvitz {\rm and} Courant} {\cite{Hur}}, {\it Mumford} {\cite{Mum}},
{\it Fay} {\cite{Fay}}, {\it Griffiths {\rm and} Harris} {\cite{Grif}}
{\it Belokolos et al} {\cite{bim}} {\it Igusa} {\cite{Igusa}}.

Notations: ${\bf C}$ - complex plane;
$\bar{\bf C}$ - compactified complex plane (Riemann sphere); ${\bf CP}^n$ -
the complex projective space of dimension $n$.
 The ends of the proofs are marked by $\Box$.

\section{Compact Riemann surfaces}

\subsection{Algebraic curves (compact Riemann surfaces)}

Let's start from the following

\begin{definition}
Algebraic curve ${\cal L}$ is a submanifold in projective space
${\bf CP}^2$ defined by the equation $R(\omega,\lambda)=0$,
where
$(\omega,\lambda)$ are local coordinates in ${\bf CP}^2$;
 $R$ is an irreducible polynomial in $\omega$ and $\lambda$.
Curve ${\cal L}$ is called non-singular if the complex gradient of the
function $R(\omega,\lambda)$  does not vanish i.e.

\begin{equation}
\nabla R\equiv (\frac{\partial R}{\partial \omega},\frac{\partial R}
{\partial \lambda})\neq 0
\label{ns}
\end{equation}
\end{definition}

Further in this chapter we shall consider only non-singular
algebraic curves.

According to the classical 
point of view an algebraic curve is 
represented as a covering of the 
complex plane. Let polynomial
$R$ have degree $m$ in $\omega$. Then the equation of curve ${\cal L}$
may be written as follows:

\begin{equation}
a_0(\lambda)\omega^m + a_1(\lambda)\omega^{m-1}+...+a_m (\lambda) = 0
\label{pol}
\end{equation}
where $a_0(\lambda),...,a_m(\lambda)$ are some polynomials and
$a_0(\lambda)$ is not identically zero. So for every $\lambda$  in general case we have $m$ corresponding values of $\omega$; therefore,
curve ${\cal L}$ may be considered as $m$-sheeted covering of $\lambda$-plane.
However, for some values of $\lambda$ equation (\ref{pol}) may have less
then $m$ different roots; the points of this kind are called the
"branch points". Apparently this definition is equivalent to the following
more exact

\begin{definition}
Point $P$ is called the branch point (in $\lambda$-plane) iff

\[ \frac{\partial R}{\partial \omega}(P) = 0 \]

\end{definition}

System of two equations $R(\omega,\lambda)=0$
and $\frac{\partial R}{\partial \omega}(\omega,\lambda)=0$
has in non-trivial situation a finite number of solutions, and, therefore,
there is a finite set of the branch points $P_1,...,P_N$.

Of course, the definition of the branch point is non-invariant under
the interchange $\omega\leftrightarrow\lambda$. Namely, if we want to
realize ${\cal L}$ as a covering of $\omega$-plane (having, in general,
another number of sheets) then the branch points are defined by the
equation $R_{\lambda}(\omega,\lambda)=0$. From the non-singularity
condition (\ref{ns}) we see that the system
$R_{\lambda}(\omega,\lambda)=R_{\omega}(\omega,\lambda)=0$ is
incompatible; hence the branch points in $\lambda$ - plane
can not coincide with the branch points in $\omega$ - plane.

It is convenient to draw  ${\cal L}$ as a covering of
$\lambda$ - plane in the form of the Hurvitz diagram (Fig.1) where
the horizontal lines denote the sheets (copies of $\lambda$ - plane)
and the vertical lines denote the positions of the
branch points where some sheets are
glued together.
\vskip 8.0cm
\begin{center}
{\bf Figure 1}
\end{center}
\vskip0.5cm
Another point of view on the algebraic curves is provided by the following

\begin{theorem}
An arbitrary non-singular algebraic curve ${\cal L}$ is a compact
complex-analytic manifold of dimension 1.
\end{theorem}

{\it Proof.} We have to establish the map of the
neighbourhood of the every point on ${\cal L}$ on some domain of the
complex plane and to show that the resulting transition functions
from ${\bf C}$ to ${\bf C}$ are complex-analytic.
It is convenient to find for every point $P_0\in {\cal L}$
the so-called {\it local parameter} at $P_0$ the function $\tau [P_0](P)$
mapping the neighbourhood of $P_0$ into the complex plane
such that $\tau [P_0](P_0)=0$.

Let's use the realization of ${\cal L}$ as a covering of $\lambda$-plane.
If point $P_0=(\omega_0,\lambda_0)$ doesn't coincide with the branch
points i.e. $\frac{\partial R}{\partial \omega}(\omega_0,\lambda_0)\neq0$
then we can simply take

\[ \tau [P_0](\omega,\lambda)=\lambda-\lambda_0 \]
In the neighbourhood of some branch point $P_j=(\omega_j,\lambda_j)$
the situation is slightly more complicated. As we discussed above, the
non-singularity condition (\ref{ns}) implies that the point $P_j$ can
not be a branch point in $\lambda$-plane and in $\omega$-plane
realizations simultaneously.
Therefore, in the neighbourhood of $P_j$ we can define the
local parameter as follows:

\begin{equation}
\tilde{\tau} [P_j](\omega,\lambda)=\omega-\omega_j
\label{lp}
\end{equation}

To find the local parameter at $P_j$ in terms of variable $\lambda$ we
can notice that condition $\frac{\partial R}{\partial \omega}
(\omega_j,\lambda_j)=0$ immediately implies together with equation
$R(\omega,\lambda)=0$ that

\[ \lambda-\lambda_j=(\omega-\omega_j)^{k_j+1}(C_j+o(1)),\;\;\;\;\;
\lambda\rightarrow \lambda_j \]
for some $k_j>0$ and constant $C_j$. Power $k_j+1$ is obviously equal to
the number of the sheets of ${\cal L}$ glued at $P_j$. Number $k_j$ is called
the degree of the branch point $P_j$.

So instead of (\ref{lp}) we can choose in the neighbourhood of $P_j$
the alternative local parameter

\[ \tau  [P_j](\omega,\lambda)=(\lambda-\lambda_j)^{\frac{1}{k_j+1}} \]

To provide compactness we have to find the local parameter in the neighbourhood of the
infinite point in $\lambda$-plane. Obviously, if $\infty$ is not a branch
point, we can choose

\[ \tau [\infty](\omega,\lambda)=\lambda^{-1}; \]
if $\lambda=\infty$ is a branch point of degree $k$ then

\[ \tau [\infty](\omega,\lambda)= \lambda^{-\frac{1}{k+1}} \]

Finally, it is easy to see that all coordinate maps described above
are compatible one to another i.e. that related transition functions are
holomorphic. $\Box$

The converse statement is also true  i.e. an arbitrary compact one-dimensional
complex manifold is conformally equivalent to some algebraic curve
{\cite{Grif}}.

As a corollary of the theorem 1 we obtain that ${\cal L}$ is a compact
oriented topological manifold of real dimension 2. So applying the
following

\begin{theorem}
Any compact oriented manifold of real dimension 2 is topologically
equivalent to sphere with finite number of handles.
\end{theorem}
(proof see in {\cite{Spring}})
one obtains that topologically the algebraic curve represents the sphere
with handles. The number of handles is called the genus; in the sequel we
denote it by $g$.

Curves with $g=1$ called elliptic are topologically equivalent to torus.

The above treatment may be conveniently illustrated for hyperelliptic
curves defined by

\begin{equation}
\omega^2-\prod_{j=1}^{n}(\lambda-E_j)=0,\;\;\;\;E_i\neq E_j,\;\;\;i\neq j
\label{he1}
\end{equation}

Curve ${\cal L}$ considered as the covering of $\lambda$-plane  
consists of two sheets; points $(\omega,\lambda)$ and
$(-\omega,\lambda)$ have the same projection on $\lambda$-plane and
lie on the different sheets of ${\cal L}$. Points $E_j,\;j=1,...,n$
are obviously the branch points. If $n$ is odd $(n=2g+1)$
then the infinite point $(\lambda=\infty)$ is the branch point on ${\cal L}$
too; if $n$ is even $(n=2g+2)$ then $\lambda=\infty$ is an ordinary point
and we have on ${\cal L}$ two infinities: ${\infty^1}$ and ${\infty^2}$.
As we shall see below, the genus of curve ${\cal L}$ in both cases
($n=2g+1$ and $n=2g+2 $) is equal to $g$. The local parameter on ${\cal L}$
in the neighbourhood of every point $P_0(\omega_0,\lambda_0),\;\;\;
\lambda _0\neq E_i,\infty$ may simply be chosen as

\[ \tau [P_0]=\lambda-\lambda_0; \]
when $\lambda_0=E_i$, we can take

\[ \tau [P_0]=\sqrt{\lambda -E_i} \]
Finally, if $\infty$ is not a branch point $(n=2g+2)$ then

\[ \tau [\infty]=\frac{1}{\lambda} \]
and if $\infty$ is a branch point $(n=2g+1)$ then

\[ \tau [\infty]=\frac{1}{\sqrt{\lambda}} \]

\subsection{Canonical basis of cycles}

Consider the first homologic group $H_1({\cal L},{\bf Z})$.
For arbitrary two elements of this group - two closed cycles $a$ and $b$ -
we can define an integer number - the intersection index $a\circ b$. If $a$
and $b$ cross each other in one point as shown in Fig.2a
then $a\circ b=-b\circ a=1$; in Fig.2b  $a\circ b=-1$.
\vskip7.0cm
\begin{center}
{\bf Figure 2}
\end{center}
\vskip0.5cm
 More
rigorously, if contours $a$ and $b$ cross in one point $P$ and the basis
in the tangent plane at $P$ consisting from their tangent vectors is
positively oriented then $a\circ b=1$ and $a\circ b=-1$ in the opposite case
(here we fix the positive orientation of curve ${\cal L}$).
If contours $a$ and $b$ cross each other in the several points
$P_1,...,P_N$ then $a\circ b$ is defined as a sum of the intersection indexes
at $P_i$ (see, for example, Fig.2c). As a result we obtain a bilinear map

\[ H^1({\cal L},{\bf Z})\times H^1({\cal L},{\bf Z}) \rightarrow {\bf Z} \]

If the genus of ${\cal L}$ is equal to $g$ then it is always possible
{\cite{Spring,Grif}} to choose in $H_1({\cal L},{\bf Z})$ the canonical
basis of $2g$ cycles

\[ a_1,...,a_g,b_1,...,b_g \]
having the following matrix of intersections

\begin{equation}
\begin{array}{cc}
a_i\circ a_j=b_i\circ b_j=0 \\
a_i\circ b_j=\delta_{ij}
\end{array}
\label{cb}
\end{equation}

Examples of the canonical basis on an abstract curve of genus 2 and on the
hyperelliptic curve
\begin{equation}
\omega^2=\prod_{i=1}^{2g+2}(\lambda-E_i)
\label{he}
\end{equation}
are presented in Fig.3a,b  .
\vskip8.0cm
\begin{center}
{\bf Figure 3: a {\rm and} b }
\end{center}
\vskip0.5cm

The choice of the canonical basis is not unique. Using the definition
(\ref{cb}) it is easy to verify that if the basis $(a_j,b_j)$ is canonical
then the basis
\begin{equation}
\tilde{a}_i=\sum_{j=1}^{g}(P_{ij}a_{j}+S_{ij}b_{j}),\;\;\;\;\;
\tilde{b}_i=\sum_{j=1}^{g}(Q_{ij}a_{j}+T_{ij}b_{j})
\label{nb}
\end{equation}
is canonical too iff the $2g\times 2g$ integer-valued matrix

\[ M\,=\,\left(\begin{array}{cc} P\;\;\;S \\ Q\;\;\;T \end{array}\right) \]
is symplectic i.e.
\[ MJM^{t}=J,\;\;\;\;
J=\left(\begin{array}{cc} 0\;\;\;I \\ -I\;\;\;0 \end{array}\right) \]
where $I$ is $g\times g$ unit matrix, $M^t$ denotes the matrix
transposed to $M$.

\section{Abelian differentials}

\subsection{Holomorphic differentials}

\begin{definition}
Differential 1-form $dU(P)$ on algebraic curve ${\cal L}$ is called the
holomorphic differential (or abelian differential of the first kind)
if in the neighbourhood of every point $Q\in{\cal L}$ it may be
represented as follows:
\[ dU=f(\tau)d\tau,\;\;\;\;\;P\sim Q \]
where $\tau$ is the local parameter at $Q$, $f(\tau)$ is a holomorphic
function.
\end{definition}

Obviously all holomorphic differentials on ${\cal L}$ constitute the
linear vector space.

\begin{theorem}
The dimension of the vector space of the holomorphic differentials on
${\cal L}$ coincides with its genus $g$.
\end{theorem}

As we shall prove below by the analysis of the cyclic periods of the
holomorphic differentials, this dimension is always less than or equal to $g$.
For the proof of existence of $g$ linearly independent holomorphic
differentials on an arbitrary genus $g$ algebraic curve
see {\cite{Spring,Hur}}. In the case of hyperelliptic
Riemann surface (\ref{he1}) (both for $n=2g+1$ and $n=2g+2$ cases) these
differentials may be written out explicitly:
\begin{equation}
dU_k(P)=\frac{\lambda^{k-1} d\lambda}{\omega},\;\;\;\;k=1,...,g
\label{hhe}
\end{equation}
The holomorphicity of the differentials (\ref{hhe}) on ${\cal L}$ may
be easily verified using the explicit form of the local parameter
at the different points of ${\cal L}$.

It is easy to see that an arbitrary holomorphic differential $dU(P)$
is closed: in the neighbourhood of the every point on ${\cal L}$ we
have
\[ d(f(\tau)d\tau) = \frac{\partial f(\tau)}{\partial \bar{\tau}}
d\bar{\tau}\wedge d\tau-\frac{\partial f(\tau)}{\partial \tau}
d\tau\wedge d\tau = 0 \]

An arbitrary antiholomorphic differential is, of course, closed too.

Taking into account the
obvious relation $\oint_{C} dW=0$, where $C$ is an arbitrary closed
cycle homological to zero, we can correctly define the cyclic periods
of an arbitrary closed differential $dW(P)$ by
\[ A_j \equiv \oint_{a_j} dW(P),\;\;\;\;\;B_j\equiv \oint_{b_j}dW(P) \]

To correctly define the function  $\int_{P_0}^{P} dW$, where $P_0$ is some
fixed point, we can make the following. Let all basic cycles
$a_j,b_j, \;j=1,...,g$ pass through some fixed point $Q\in {\cal L}$
(see Fig.4a for $g=2$). Then, cutting ${\cal L}$ along all these cycles
we obtain the simply-connected domain $\tilde{\cal L}$ - the so-called
fundamental polygon having $4g$ sides (Fig.4b) (this is in fact the
fundamental domain of related to ${\cal L}$ Fuchsian group {\cite{Nag}}).
\vskip8.0cm
\begin{center}
{\bf Figure 4: a} and {\bf b}
\end{center}
\vskip0.5cm 
Function
\[ F(P)=\int_{P_0}^{P}dW \]
is obviously  singlevalued  on $\tilde{\cal L}$.

The boundary of $\tilde{\cal L}$ may be represented as follows
\begin{equation}
\partial\tilde{\cal L} = \sum_{j=1}^{g}(a_j^+ + b_j^+ - a_j^- -b_j^-)
\label{bfp}
\end{equation}

Let's prove the following 

\begin{lemma}
Let $dW(P)$ and $dW'(P)$ be two closed differentials on ${\cal L}$ and
$A_j,B_j,A'_j,B'_j,\,j=1,...,g$ be sets of related cyclic periods;
$W(P)=\int_{P_0}^{P} dW$ (where $P_0\in\tilde{\cal L}$ is some fixed point)
is a function on $\tilde{\cal L}$. Then
\begin{equation}
\int\int_{\cal L} dW\wedge dW' \,=\,\oint_{\partial \tilde{\cal L}}
W(P)dW'(P)=\sum_{j=1}^{g} (A_j B'_j - A'_j B_j)
\label{lem}
\end{equation}
\end{lemma}
(we shall use this statement in two situations: when $dW$ and $dW'$ are
both holomorphic or one holomorphic and one anti-holomorphic)

{\it Proof.} The first equality in (\ref{lem}) is a simple corollary
of the Stokes formula and the requirement that differential $dW'$ is
closed. To prove the second one  use the
representation (\ref{bfp}) of $\partial \tilde{\cal L}$:

\[ \oint_{\partial \tilde{\cal L}} W(P)dW'(P) =\]
\[ \sum_{j=1}^{g}
\left(\int_{a_j^+}-\int_{a_j^-}\right)W(P)dW'(P) +
\sum_{j=1}^{g}\left(\int_{b_j^+} -\int_{b_j^-}\right)W(P)dW'(P) \]
Now, for every $j$th part of the boundary of the 
fundamental polygon consisting of
four sides $a_j^+,b_j^+,a_j^-,b_j^-$ (Fig.5) 
\vskip6.0cm
\begin{center}
{\bf Figure 5}
\end{center}
\vskip0.5cm
we can easily see that
for the coinciding on ${\cal L}$ points $P_j^{\pm}$ lying on the cycles $a_j^{\pm}$
respectively
\[ W(P_j^+)-W(P_j^-) = -B_j \]
and for the points $Q_j^{\pm}$ lying on cycles $b_j^{\pm}$
\[ W(Q_j^+)-W(Q_j^-) = A_j \]

The differential $dW'(P)$ is singlevalued on ${\cal L}$ and, therefore, is
the same on the different banks of $a_j$ and $b_j$. As a result we have
\[ \oint _{\partial \tilde{\cal L}} W(P)dW'(P) =
\sum_{j=1}^{g}(-B_j)\oint_{a_j}dW'(P) + \sum_{j=1}^{g} A_j
\oint_{b_j} dW'(P) =\]
\[= \sum _{j=1}^{g} (A_j B'_j-B_j A'_j ) \;\;\;\Box\]

>From this basic lemma we can derive several  corollaries.

\begin{corollary}
Let $dU(P)$ be a non-zero holomorphic differential on ${\cal L}$ with
cyclic periods
$A_j, B_j,\,j=1,...,g$. Then

\begin{equation}
{\rm Im}\sum_{j=1}^{g} A_j\bar{B}_j < 0
\label{c1}
\end{equation}
\end{corollary}

{\it Proof.} Let's choose in lemma 1 $dW(P)= dU(P)$,
$dW'(P)= d\bar{U}(P)$. Then $A'_j=\bar{A}_j$, $B'_j=\bar{B}_j$;
besides that, at every point of ${\cal L}$ we can choose the local
parameter $\tau=x+iy$ in such a way that $dU=d\tau=dx+idy$ and
\[ dU\wedge d\bar{U}=(dx+idy)\wedge (dx-idy) = -2i dx\wedge dy \]

Taking into account that the curve ${\cal L}$ is positively-oriented,
we obtain
\[ {\rm Im}\int\int_{\cal L} dU\wedge d\bar{U} < 0 \]
that immediately entails (\ref{c1}). $\Box$

>From (\ref{c1}) we obtain the following

\begin{corollary}
If all the 
periods of the holomorphic differential $dU(P)$ are equal to zero then
$dU(P)\equiv 0$
\end{corollary}

This corollary obviously implies that the dimension of the linear space
of the holomorphic differentials on ${\cal L}$ is  equial to $\leq g$
(otherwise we would be able to construct a non-trivial holomorphic
differential with zero periods). Together with the proof of existence
of $g$ linearly independent holomorphic differentials that may be found
in {\cite{Spring,Hur}} (for hyperelliptic curve (\ref{he}) they are
given by the exact formulae (\ref{hhe})) this gives Theorem 3.

Let $dW_1,...dW_g$ be linearly independent holomorphic differentials
on ${\cal L}$. Then their matrix of $a$-periods
\[ A_{ij} = \oint_{a_i} dW_j \]
is obviously non-degenerated (otherwise some non-trivial linear
combination of them has all zero $a$-periods that contradicts  Corollary 2).
So it is possible to choose another basis of holomorphic differentials
$dU_j(P),\;j=1,...,g$ normalized by the condition
\begin{equation}
\oint_{a_k} dU_j(P) = \delta_{kj}
\label{nc}
\end{equation}

Differentials $dU_j$ constitute the {\it canonical basis of holomorphic
1-forms} dual to the canonical basis of cycles $(a_j,b_j)$.

Now introduce the new fundamental object
associated to the curve ${\cal L}$ - the {\it matrix of b-periods}

\begin{equation}
B_{jk} = \oint_{b_j} dU_k
\label{bper}
\end{equation}

\begin{theorem}
The matrix $B$ is symmetric and has positively-defined imaginary part.
\end{theorem}

{\it Proof.} To prove that $B$ is symmetric let's put in lemma 1
$dW=dU_j$, $dW'=dU_k$; then obviously $dU_j\wedge dU_k = 0$ and the r.h.s.
of (\ref{lem}) gives $B_{kj}=B_{jk}$.

To prove the second part consider for an arbitrary $x\in {\bf R}^g$
the linear combination
\[ dU = \sum_{j=1}^{g} x_j dU_j ;\]
its $a-$ and $b-$periods are the following:
\[ A_k=x_k;\;\;\;\;\;B_k=\sum_{j=1}^{g} x_j B_{kj} \,. \]
Applying the Corollary 1 to the differential $dU$ we obtain
\[ {\rm Im}\sum_{j,k=1}^{g}x_j \bar{B}_{kj} x_k = {\ Im} \sum_{k=1}^{g} A_k \bar{B}_k
< 0 \]
and, therefore, ${\rm Im}\langle Bx,x\rangle$ is always positive
($\langle .,.\rangle$ is the ordinary scalar product).$\Box$

Matrix $B$ is a very important object characterizing the curve ${\cal L}$.
The famous Shottki problem is the problem of isolating 
the matrices of $b$-periods of algebraic curves 
among the whole family of symmetric matrices
with positively-defined imaginary part. The essential progress in this
field was recently achieved in the framework of algebro-geometrical
 approach to the
integrable equations {\cite{Shiota}}.

It is easy to verify that if we choose the new basis of cycles
$(\tilde{a}_j,\tilde{b}_j)$ related to basis $(a_j,b_j)$ by (\ref{nb})
then the corresponding matrices of $b$-periods $\tilde{B}$
and $B$ are related as follows:
\begin{equation}
\tilde{B}=(TB+Q)(SB+P)^{-1}
\label{nm}
\end{equation}

Now we can construct the so-called {\it Jacobi manifold} of curve ${\cal L}$:

\begin{definition}
The complex torus of dimension $g$ $J({\cal L})$ defined as the following
quotient

\[ J({\cal L})={\bf C}^{g}/\{N+BM\}\;\;, \]
where $N,M\in{\bf Z}^{g}$, is called the Jacobi manifold (or Jacobian) of the
curve ${\cal L}$.
\end{definition}

The transformation of the matrix $B$ (\ref{nm}) with respect to
 the change of the
canonical basis of cycles leads to some non-essential transformations
of $J({\cal L})$ (see {\cite{dub,Mum}}).

Next consider the meromorphic 1-forms on ${\cal L}$.

\subsection{Meromorphic differentials}

\begin{definition}
Meromorphic 1-form having on ${\cal L}$ the unique pole at point $Q$
with the following local expansion:
\[ (\frac{1}{\tau^{n+1}}+O(1))d\tau\;\;, \]
where $\tau$ is the local parameter in the 
neighbourhood of $Q$ and $n\in {\bf N}$, is called the
Abelian differential of the 2nd kind.
\end{definition}

\begin{definition}
Meromorphic 1-firm having on ${\cal L}$ the simple poles at
$P=Q$ and $P=R$ with the residues $+1$ and $-1$ respectively is called
the Abelian differential of the 3rd kind.
\end{definition}

For an arbitrary Abelian differential $dW$ on ${\cal L}$ (of 1st, 2nd, or
3rd kind) related (in general case non-singlevalued on ${\cal L}$) function
\[ W(P)=\int_{P_0}^{P} dW\;\;, \]
where $P_0\in{\cal L}$ is some fixed point, is called the {\it Abelian
integral} (of 1st, 2nd or 3rd kind respectively).

Notice that the Abelian integrals (or simply integrals)
of the 1st kind are everywhere
holomorphic; integrals of the 2nd kind are meromorphic and the integrals of
the 3rd kind have logarithmic singularities at $Q$ and $R$. The integrals
of the 1st and 2nd kind are singlevalued on the fundamental polygon
$\tilde{\cal L}$; integral $W_{QR}$ is singlevalued on
$\tilde{\cal L}$ with the cut $[Q,R]$.

It appears possible to prove that the meromorphic differentials
of the 2nd and 3rd kind exist for arbitrary positions of the
poles and arbitrary related singular parts {\cite{Spring,Hur}}
(for differentials of the third kind the only restriction is the
vanishing of the sum of all the residues). 

For hyperelliptic curve (\ref{he}) these
integrals may be easily written out explicitly.
Namely, if $Q=(\omega_1,\lambda_1)$ and $R=(\omega_2,\lambda_2)$
don't coincide with the branch
points, we can express the related differentials of the 3rd kind as
follows:
\begin{equation}
dW=\frac{d\lambda}{2\omega}\left(\frac{\omega+\omega_1}{\lambda-\lambda_1}
-\frac{\omega+\omega_2}{\lambda-\lambda_2}\right)
\label{md3}
\end{equation}

For the differential of the 2nd kind having pole at $Q$ with $n=1$ we
have:
\begin{equation}
dW=\frac{d\lambda}{2\omega}\left(\frac{\omega+\omega_1}{(\lambda-\lambda_1)^2}
+\frac{\omega'(\lambda_1)}{\lambda-\lambda_1}\right)
\label{md2}
\end{equation}
(it is not difficult to write analogous expressions if $\lambda_1$ or
$\lambda_2$ coincide with the branch points).

Of course, the meromorphic differential having prescribed poles and the
singular parts is not unique: it is defined up to an arbitrary holomorphic
differential. To get rid of this non-uniqueness we can 
normalize meromorphic differentials as follows

\begin{definition}
Abelian differential of the 2nd or the 3rd kind is called normalized if
all its $a$-periods are equal to zero.
\end{definition}

An arbitrary meromorphic differential may obviously be normalized by
adding some linear combination of basic holomorphic differentials.
The normalization together with the positions of the poles and 
values of related
singular parts defines the meromorphic differential uniquely.
(if $dW_1$ and $dW_2$ are two normalized holomorphic differentials with
the same poles and the singular parts then $dW_1-dW_2$ is a holomorphic
differential with vanishing $a$-periods, and, therefore, $dW_1=dW_2$).

The $b$-periods of normalized meromorphic differentials 
may be easily expressed in terms of holomorphic differentials. Namely,

\begin{lemma}
The $b$-periods of the normalized abelian differentials of the 2nd and
the 3rd kind may be represented in the following form:

\begin{equation}
(B_Q^{(n)})_j \equiv\oint_{b_j} dW_Q^{(n)} = \frac{2\pi i}{n!}
\frac{d^{(n-1)} f_j}{d\tau ^{(n-1)}}(\tau=0)
\label{b2}
\end{equation}
where function $f_j$ describes the local behaviour of holomorphic
differential $dU_j$ at the 
\[ dU_j(P)=f_j(\tau)d\tau \]
($\tau$ is the local parameter at $Q$);

\begin{equation}
(B_{QR})_j\equiv\oint_{b_j} dW_{QR} = 2\pi i (U_j(Q)-U_j(R))
\label{b3}
\end{equation}
\end{lemma}
{\it Proof.} Consider (\ref{b3}) ((\ref{b2}) may be proved
in the same way). In analogy to the proof of lemma 1 we have:
\[ (\ast)\equiv\oint_{\partial\tilde{\cal L}} U_j(P)dW_{QR}(P) =
\sum_{k=1}^{g}(A_{jk}(B_{QR})_k - (A_{QR})_k B_{jk}) = (B_{QR})_j \]
where $A_{jk}=\delta_{jk}$ and $B_{jk}$ are $a$- and $b-$ periods
of basic holomorphic differential $dU_j(P)$; $(A_{QR})_k =0$ are
$a$-periods of $dW_{QR}^{(n)}$. Calculating $(\ast)$ in another way
according to residue theorem, we obtain (\ref{b3}).

\section{Meromorphic functions}

\subsection{Meromorphic functions. Abel theorem}

Now we are in position to start the treatment of the meromorphic
functions on ${\cal L}$. Any meromorphic function may  be
considered as some linear combination of the abelian integrals of the
2nd kind which has vanishing cyclic periods. 
This condition appears very crucial:
the space of meromorphic functions has much more complicated structure
then the space of the abelian integrals. This structure is simplest for
${\cal L}=\bar{\bf C}$ (Riemann sphere). Here every meromorphic function
may be represented as the ratio of two polynomials:
\begin{equation}
 f(\lambda)=C\frac{(\lambda-a_1)...(\lambda-a_n)}{(\lambda-b_1)...
(\lambda-b_m)}
\label{rf}
\end{equation}

Notice the difference between the theory of functions on
the ordinary non-compact complex plane ${\bf C}$ and its
compactification $\bar{\bf C}$. We can claim, for example, that every
holomorphic function on $\bar{\bf C}$ is a constant; the same statement
for ${\cal C}$ is not true (counterexample is the function $\exp\lambda$).
Moreover, on $\bar{\cal C}$ function (\ref{rf}) has the same number
of poles and zeros (taking into account their order); on ${\bf C}$
this is, generally speaking, not true.

Some features of the space of meromorphic functions on compact Riemann
surface of non-zero genus appear very similar to $\bar{\bf C}$ 
case (for example, the statement that  every holomorphic function 
on algebraic curve is a constant, or the fact that the number of poles
is always equal to the number of zeros). However, in other aspects the
genus plays a crucial role. For example, if $g>0$, then the positions
of zeros and poles of the meromorphic function should obey some
additional restrictions (in contrast to $g=0$, where the poles and zeros
in (\ref{rf}) may be defined arbitrarily).

Now let's prove some facts about the meromorphic functions on
compact Riemann surfaces.

\begin{theorem}
The number of poles of meromorphic non-constant function
$f(P)$ on ${\cal L}$ is always finite
and equal to the number of zeros. (We calculate the number of poles and zeros taking into account their order.)
\end{theorem}

{\it Proof.} The finiteness of the number of zeros obviously follows from the
compactness of ${\cal L}$ and the existence of  non-trivial
Taylor expansion of $f(P)$ at  every point.

To prove that the number of zeros of $f(P)$ is equal to the number of poles
it is sufficient to integrate expression $d\log f(P)$ along the boundary
of fundamental polygon. On one hand, it vainishes since the cyclic
periods along the basic cycles of this differential are equal to zero.
On  the other hand, it is equal to the difference between the number of
zeros and the number of poles of $f(P)$ (up to a factor $2\pi i$). $\Box$

Now it is natural to ask when we can construct function $f(P)$
having prescribed set of zeros and poles. To  answer this
question define the {\it Abel map} $U$ from curve ${\cal L}$ into its
Jacobi manifold:
\[ U:{\cal L}\rightarrow J({\cal L}) \]
in the following way:
\begin{equation}
U_j(P)=\int_{P_0}^{P} dU_j\;,\;\;\;\;j=1,...,g
\label{Amap}
\end{equation}
where $dU_j(P)$ are normalized basic holomorphic differentials; $P_0\in
{\cal L}$ is some fixed point and the path from $P_0$ to $P$ is chosen
in the same way for all $j$. To prove that definition
(\ref{Amap}) is correct we have to
 check its  independence of the choice of
the path between $P_0$ and $P$. Indeed, if we add to this  path some
linear combination of basic cycles
\[ \gamma=\sum_{k} n_ka_k +\sum_{k}m_kb_k \]
then in the r.h.s. of (\ref{Amap}) we get the additional term
\[ \oint_{\gamma} dU_j = n_j +\sum_{k} B_{jk} m_k \]
i.e. the vector of the lattice defining $J({\cal L})$; so the change of
the path in (\ref{Amap}) doesn't change the position of the point on
$J({\cal L})$ and (\ref{Amap}) correctly defines the map from ${\cal L}$
to $J({\cal L})$.

Now we can formulate the following
\begin{theorem}[Abel]
Two sets of points: $P_1,...,P_n$ and $Q_1,...,Q_n$ on ${\cal L}$ are the 
sets of
poles and zeros (respectively)
of some meromorphic function iff
\begin{equation}
\sum_{j=1}^{g} (U(P_j)-U(Q_j)) \equiv 0
\label{At}
\end{equation}
(symbol $\equiv$ here and below means that the left side and the right side
define
the same point on $J({\cal L})$ i.e. differ by some vector of the lattice
$\{m+Bn\}$).
\end{theorem}

{\it Proof.}
Let meromorphic function $f(P)$ have poles and zeros at
$P_1,...,P_n$ and $Q_1,...,Q_n$ respectively. Then the meromorphic
differential $d\Omega(P)=d\log f(P)$ has simple poles at the points
$P_1,...,P_n$ with  the residues equal to $-1$ and at the points
$Q_1,...,Q_n$ with the residues equal to
$+1$; therefore, it may be represented as follows:
\[ d\Omega(P)=\sum_{j=k}^{n} dW_{Q_k P_k}(P) +\sum_{k=1}^{n}
C_k dU_k(P) \]
where $dW_{Q_k P_k} (P)$ are normalized abelian differentials of the
3rd kind, $dU_k(P)$ are basic holomorphic differentials and $C_k$ are
some constants.

>From the singlevaluedness of function $f(P)$ on ${\cal L}$ we conclude
that the integral of $d\Omega$ along an arbitrary closed contour
should be equal to $2\pi i n$ for some integer $n$. In particular,
\[ \oint_{a_j} d\Omega =2\pi i n_j,\;\;\;\;\;\;
\oint_{b_j} d\Omega = 2\pi i m_j \]
where $n_j,m_j \in {\bf Z}$. So, taking into account the
normalization condition (\ref{nc}) and formulae for the $b$-periods of
differentials $dW_{Q_jP_j}$ (\ref{b3}), we obtain
\[ 2\pi i n_j = C_j \]
and
\[ 2\pi i m_j = 2\pi i \sum_{k=1}^{g} (U_j(P_k)-U_j(Q_k)) + 2\pi i
\sum_{k=1}^{g} B_{jk} n_k \]
Therefore
\[ \sum_{k=1}^{g} (U_j(P_k)-U_j(Q_k)) = m_j - \sum_{k=1}^{g} B_{jk} n_k \]
that yields (\ref{At}).

Conversely, having a set of the points $P_1,...,P_n$ and $Q_1,...,Q_n$
obeying (\ref{At}), we can construct the meromorphic differential
\[ d\Omega=\sum_{k=1}^{g} dW_{P_kQ_k} + 2\pi i \sum_{k=1}^{g} n_k dU_k \; \]
Then the function
\[ f(P)=\exp\int_{P_0}^{P}  d\Omega  \]
has poles at $P_1,...,P_n$, zeros at $Q_1,...,Q_n$ and is singlevalued on
${\cal L}$ for some integer $n_k,\;k=1,...,g$ due to the normalization
conditions of $dW_{P_kQ_k}$ and relations (\ref{At}). $\Box$

Notice that the Abel theorem doesn't give us the  effective criterion for
the existence of meromorphic functions with prescribed zeros and poles.
To formulate the further results in this direction it is convenient to
define some new objects.

\subsection{Divisors on algebraic curves}

\begin{definition}
The formal linear combination
\[ D=\sum_{j=1}^{N} n_j P_j\;, \]
where $P_j\in {\cal L},\;n_j\in {\bf Z},\,j=1,...,N$, is called a divisor
on the curve ${\cal L}$.
\end{definition}

All divisors on ${\cal L}$ constitute the abelian group with
respect to the obvious summation operation: for
\[ D=\sum_{j=1}^{N}n_j P_j\,\;\;\;{\rm and}\;\;\;\;\;
D'=\sum_{j=1}^{N'} n'_j P'_j \]
we define
\[ D+D'=\sum_{j=1}^{N} n_j P_j + \sum_{j=1}^{N'} n'_j P'_j \]
Number
\[ \deg D =\sum_{j=1}^{N} n_j \]
is called the {\it degree} of the divisor $D$.

For an arbitrary meromorphic function $f(P)$ having zeros at
the points $Q_1,...,Q_n$ of the order $q_1,...,q_n$ and poles at 
the points $P_1,...,P_m$
of the order $p_1,...,p_m$ respectively, the divisor
\[ (f)=p_1 P_1+...+p_m P_m - q_1 Q_1 -...-q_n Q_n \]
is called the divisor of function $f$. Theorem 5 shows that
 $\deg(f)=0$.

Two divisors $D$ and $D'$ are called linearly equivalent if divisor
$D-D'$ is a divisor of some meromorphic function. Obviously in this case
$\deg D=\deg D'$. The divisor of arbitrary meromorphic function is
linearly equivalent to zero.

We can associate  the divisor of
zeros and poles $(dW)$
to any meromorphic differential $dW$ 
(the zeros and poles of $dW$ are defined as zeros
and poles of related local function $h(\tau)$ in the local
representation $dW=h(\tau)d\tau$ at the every point of ${\cal L}$).

It is easy to prove that the divisors $(dW)$ and $(dW')$ of arbitrary
two meromorphic differentials $dW$ and $dW'$ are linearly equivalent:
\[ (dW)-(dW')=(\frac{dW}{dW'}) \]
where $\frac{dW}{dW'}(P)$ is a meromorphic function on ${\cal L}$.

The equivalence class of the divisors of meromorphic differentials
is called the {\it canonical class}
of of curve ${\cal L}$ and is defined by $C$.

Now define the Abel map on the divisors by linearity: for
\[ D=\sum_{j=1}^{N} n_i P_i \]
we put
\[ U(D) = \sum_{j=1}^{N} n_j U(P_j) \]

The language of divisors is very convenient to formulate the theorems
related to meromorphic functions on ${\cal L}$. For example, the Abel
theorem may be formulated as follows:

{\it Divisors $D$ and $D'$ are linearly equivalent iff}

\[ 1.\deg D =\deg D' \]

\[ 2.\;U(D)\equiv U(D') \]

(The item 2. means that $U(D)$ and $U(D')$ define the same point of
$J({\cal L})$)

Divisor $D=\sum_{j=1}^{N} n_j P_j$ is called {\it positive} if $n_j >0$
for all $j=1,...,N$.

Now we can define the partial ordering
on the set of the divisors assuming $D\geq D'$ iff $D-D'$ is a
positive divisor.

With  every divisor $D$ we can associate the linear vector space
$L(D)$ of meromorphic functions $f(P)$ on ${\cal L}$ obeying the
condition
\begin{equation}
(f)\geq -D
\label{LD}
\end{equation}
In less abstract language it means that given any divisor
$D=n_1 P_1 +...+ n_k P_k -m_1 Q_1 - ...- m_l Q_l\;\;(n_j,m_j>0)$,
function $f(P)$ belongs to the space $L(D)$ iff

1. It has no poles on ${\cal L}$ outside $P_1,...,P_k$ and the order
of its pole at $P_j$ is  $\leq n_j$, $j=1,...,k$.

2. It has zeros at $Q_j$ of the order at least $m_j,\;j=1,...,l$.

The dimension of the linear space $L(D)$ is denoted by $l(D)$. If
divisors $D$ and $D'$ are linearly equivalent then apparently
$l(D)=l(D')$ (if $D$ is linearly equivalent to $D'$ then the isomorphism
between the spaces $L(D)$ and $L(D')$ may be established by the
multiplication of any function from $L(D)$ 
on the function $h(P)$ obeying condition $(h)= D-D'$).

Below we present some facts about the number $l(D)$. The central result
here is the classical version of the Riemann-Roch theorem establishing the
link between  $l(D)$ and $l(C-D)$ where $C$ is an arbitrary divisor from
the canonical class (divisor of arbitrary meromorphic differential
on ${\cal L}$). Our treatment here mainly follows {\cite{Hur}}.

\subsection{Riemann-Roch theorem}

Let's first prove the following 

\begin{theorem}
For any positive divisor $D$ on curve ${\cal L}$ of genus $g$
\begin{equation}
l(D)=\deg D -g+1+l(C-D)
\label{RRp}
\end{equation}
where $C$ is an arbitrary divisor from the canonical class.
\end{theorem}

{\it Proof.} For the sake of transparency
we restrict ourselves by divisors
$D$ having the form
\[ D=P_1+...+P_n \]
where $P_i\neq P_j$; then $L(D)$ is the linear space of the functions
that have no singularities on ${\cal L}$ except, probably, the simple
poles at $P_1,...,P_n$ (the generalization to higher order poles
is straightforward). Let also for definiteness $n\geq g$ (the opposite
case may be considered analogously).

Consider the set of normalized (all $a$-periods are zero) meromorphic
differentials of the 2nd kind $dW_{P_j}^{(1)} (P),\;\;j=1,...,n$.
(Existence of differentials $dW_{Q}^{(n)}$ on ${\cal L}$ for an arbitrary
$Q\in {\cal L}$ and $n\geq 1$ is proved in {\cite{Spring,Hur}}; formula
(\ref{md2}) gives this differential for $n=1$ in the case of hyperelliptic
curve (\ref{he})).

Define by $W_j(P)$ the related integrals:
\[ W_j(P)=\int_{P_0}^{P}dW_{P_j}^{(1)},\;\;\;j=1,...,n \]
and consider the linear combination
\begin{equation}
f(P)=\sum_{j=1}^{n} \alpha_j W_j(P)
\label{f}
\end{equation}
for some $\alpha_j\in {\bf C}$. Integral $f(P)$ has simple poles at
$P_1,...,P_n$ and no other singularities. Besides that, all its $a$-periods
vanish. So this is a meromorphic function from the space $L(D)$ if all its
$b$-periods also vanish. Moreover, it is obvious that any
function $f(P)\in L(D)$ may be represented in the form (\ref{f}):
$\alpha_j$ are equal the residue at $P_j$.

Defining $n$ vectors of $b$-periods
\[(B_j)_k=\oint_{b_k} dW_{P_j}^{(1)},\;\;\;\;j=1,...,n,\;k=1,...,g \]
and considering matrix $n\times g$ $M_{jk}=(B_j)_k$  we see that the
vanishing of all $b$-periods of the integral (\ref{f}) is equivalent
to the following linear system
\begin{equation}
\sum_{j=1}^{n} M_{jk} \alpha _j =0\,,\;\;\;k=1,...,g
\label{ls1}
\end{equation}

This system has $n-{\rm rank}M$ linearly independent solutions; therefore,
in $L(D)$ we obtain $n-{\rm rank}M$ linearly independent non-trivial
functions (with the different sets of residue at $P_j$); taking into
account the
constant function we obtain
\begin{equation}
l(D)=\deg D-{\rm rank}M+1
\label{lD}
\end{equation}

Now let's establish the link between  $\,{\rm rank}M$ and $l(C-D)$. It
remains to prove that
\begin{equation}
g-{\rm rank}M=l(C-D)
\label{gr}
\end{equation}

Instead of $C$ we can substitute a divisor of an arbitrary meromorphic
differential. We shall prove (\ref{gr}) in two steps. First,
show that
\begin{equation}
g-{\rm rank} M = \dim H(D)
\label{dimH1}
\end{equation}
where by $H(D)$ we denote the linear space of holomorphic
differentials having zeros at $P_1,...,P_n$.

Am arbitrary holomorphic differential $dU(P)$ on ${\cal L}$ can be represented
as a linear combination of the basic normalized differentials:
\begin{equation}
dU(P)=\sum_{k=1}^{g} \beta_k dU_k (P)
\label{dU}
\end{equation}

Now consider the following integral:
\begin{equation}
\oint_{\partial \tilde{\cal L}} W_k(P) dU(P) = (\ast),\;\;\;\;k=1,...,g
\label{ints}
\end{equation}
It is equal to zero iff $dU(P)$ vanishes at the pole of
$W_k(P)$ i.e. at $P_k$. So $dU(P)\in H(D)$ iff all the 
integrals (\ref{ints}) vanish. On the other hand, calculating
integrals (\ref{ints}) as in the proof of lemma 2.1, using normalization
conditions for $dU_k$, $W_k$ and representation (\ref{dU}) we see that
\[ (\ast) = -\sum_{j=1}^{g} \beta_j (B_k)_j \]
where as before $B_k, \,k=1,...,n$ is the vector of $b$-periods of the
integral $W_k$. So $dU(P)\in H(D)$ iff coefficients $\beta_k$
obey the following linear system:
\begin{equation}
\sum_{k=1}^{g} M_{jk} \beta_k =0,\;\;\;j=1,...,n
\label{ls2}
\end{equation}
where $n\times g$ matrix $M_{jk}$ consisting of vectors $B_k$ is the same
as in (\ref{ls1})  (notice, however, that the linear systems (\ref{ls1})
and (\ref{ls2}) are different). Since we consider the case $g\leq n$, the
system (\ref{ls2}) has $g-{\rm rank}M$ linearly independent solutions, that
implies (\ref{dimH1}).

Now to verify (\ref{gr}) it remains to prove that
\begin{equation}
\dim H(D) = l(C-D)
\label{dimH2}
\end{equation}

Let's take $C=(dW)$ where $dW$ is some meromorphic differential
($l(C-D)$ is invariant with respect to this choice). The one-to-one linear
correspondence between the linear spaces $H(D)$ and ${\cal L}((dW)-D)$ may be
established by means of differential $dW$ as follows:
\[ dU(P)=f(P) dW(P)\,,\;\;\;dU\in H(D), \;\;\;f\in  L((dW)-D) \]
So (\ref{dimH2}) is proved and, combining it with (\ref{dimH1}) and
(\ref{lD}) we obtain (\ref{RRp}). $\Box$

Relation (\ref{RRp}) allows to get information about the existence and
the number of linearly independent functions having prescribed set of the 
poles.
In particular, we have the following
\begin{corollary}[Riemann inequality]
For any positive divisor $D$ the following inequality is fulfilled:
\begin{equation}
l(D)\geq {\rm deg} D-g+1
\label{Ri}
\end{equation}
\end{corollary}

So for arbitrary $g+1$ points $P_1,...,P_{g+1}$ on ${\cal L}$ we have
$l(D)\geq 2$ and, therefore, we can always find a non-trivial
meromorphic function $f$ on ${\cal L}$ which has no singularities
except, probably, the simple poles at $P_1,...,P_{g+1}$. For the set
of $g$ poles such a function is generically absent. The divisors
$D_1+...+D_g$ for which such a function exists are called special; they
constitute the subset of complex codimension 1 in the space of
all divisors. The generic divisors are described by the following
\begin{definition}
Positive divisor is called non-special if
\[ l(D)={\rm deg} D-g+1 \;;\]
otherwise $D$ is called special.
\end{definition}

Here it is relevant to prove the following simple statement which allows
to classify the positive divisors of degree $g$ on hyperelliptic
curves:
\begin{statement}
Divisor $D=P_1+...+P_g$ with $P_j\neq P_k$ on the 
hyperelliptic algebraic curve ${\cal L}$
of genus $g$
(\ref{he}) is non-special iff points $P_1,...,P_g$ have
 different projections on $\lambda$-plane.
\end{statement}

This fact illustrates the general
situation that divisors in general position are non-special and,
therefore,
an arbitrary special divisor may be made non-special by
a "little stirring".

{\it Proof.} According to the proof of theorem 7 it is enough to
show that $\dim H(D)=0$ iff points $P_j\equiv (\omega_j,
\lambda_j)$ have different projections on $\lambda$-plane (here $H(D)$ is
the linear space of holomorphic differentials having poles at $P_1,...,P_g$).
Since an arbitrary holomorphic differential on the curve (\ref{he}) is a linear
combination of basic differentials (\ref{hhe}), it may be represented
in the form
\[ dU(P)=\frac{P_{g-1}(\lambda)}{\omega} \]
where $P_{g-1}(\lambda )$ is some polynomial of degree $g-1$. Then
conditions $dU(P_j)=0,\;j=1,...,g$ are equivalent to the linear
system
\[ P_{g-1}(\lambda_j)=0,\;\;\;j=1,...,g \] for $g$ coefficients of
polynomial $P_{g-1}$. This system has non-trivial solutions iff
its determinant that is equal simply to Vandermond determinant of
values $\lambda_1,...,\lambda_g$
\[ \Delta\equiv \prod_{j\neq k} (\lambda_j -\lambda_k) \]
vanishes  i.e. $\lambda_j =\lambda_k$ for some $j,k$. $\Box$

Another result that can be easily deduced from  theorem 7 is about the  degree
of an arbitrary divisor from the canonical class.
\begin{corollary}
For an arbitrary meromorphic differential $dW$ on ${\cal L}$ we have
\[ {\rm deg}(dW) = 2g-2 \]
\end{corollary}

{\it Proof.} As we have shown before, $\deg(dW)$ is the same for all
meromorphic differentials since $(dW)-(dW')$ is the divisor of the meromorphic
function $\frac{dW}{dW'}$. Now let's put in (\ref{RRp}) $D=(dU_0)$ where
$dU_0$ is an arbitrary holomorphic differential ($D>0$
since $dU_0$ is holomorphic). Linear space $L(D)$ consists of the functions
having poles at the points of $D$; every function $f\in L(D)$ may
obviously be represented in the form
\begin{equation}
f(P)=\frac{dU(P)}{dU_0(P)}
\label{hh}
\end{equation}
where $dU$ is an arbitrary holomorphic differential.

The linear space of meromorphic differentials on ${\cal L}$ (and, therefore,
the linear space of functions that may be represented as (\ref{hh}) for
fixed $dU_0$) has dimension $g$; therefore, $l(D)=g$. Dimension
$l(C-D)$ may be calculated
taking $C$ to be an arbitrary divisor from the
canonical class, for example, $C=(dU_0)$); then 
$l(C-D)=l(0)$ ($0$ is the zero divisor). So $l(C-D)=1$
(space $L(0)$ includes only the constant functions).
As a result we can write (\ref{RRp}) as follows
\[ g={\rm deg} D-g+1+1 \]
and, therefore, ${\rm deg} D =2g-2$ gives the degree of an 
arbitrary divisor from
the canonical class. $\Box$

Now we can prove the following {\it Riemann-Roch}
theorem (more exactly,
its "classical" version related to meromorphic functions on algebraic
curves). 

\begin{theorem}[Riemann-Roch]
The following relation is true 
for an arbitrary divisor $D$ on algebraic curve ${\cal L}$ of genus $g$
\begin{equation}
l(D)=\deg D -g+1+l(C-D)
\label{RR}
\end{equation}
\end{theorem}

{\it Proof.} As we have proved above, (\ref{RR}) is true for an arbitrary
positive divisor $D_0>0$. Let's consider the divisor
\begin{equation}
\tilde{D}=D_0+(f)
\label{DW}
\end{equation}
where $f$ is some meromorphic function. Divisors $\tilde{D}$ and
$D$ are linearly equivalent, and, therefore,
\[ \deg D_0=\deg\tilde{D}, \;\;\;l(D_0)=l(\tilde{D}),\;\;\;
l(C-D_0)=l(C-\tilde{D}) \]
So (\ref{RR}) is valid for an arbitrary divisor which may be represented as
(\ref{DW}), where $D_0$ is a positive divisor.

Now consider two cases:

1. $l(D)>0$. Take some function $f_0\in (D)$; by definition it means that
$D_0\equiv (f_0)+D>0$; so we can claim that
\[ l((f_0)+D)=\deg ((f_0)+D)-g+1+l(C-(f_0)-D) \]
or, equivalently,
\[ l(D)=\deg D -g +1 +l(C-D) \]
which coincides with (\ref{RR})

2. $l(D)=0$. Again consider two cases:

a. $l(C-D)\neq 0$ Then, taking into account item 1. and Corollary 4, we
can claim that
\[ l(C-D)=\deg (C-D)-g+1+l(D) = \]
\[ = 2g-2-\deg D -g +1 +l(D) \]
that again gives (\ref{RR}).

It remains to consider the case

b. $l(C-D)=l(D)=0$.
Here we have to prove that 
\[ \deg D =g-1 \].

Let $D=D_1-D_2$ where $D_1=P_1+...+P_k,\;\;D_2=Q_1+...+Q_l$ are two
positive divisors. Using condition $l(D)=0$ we can claim that
\[ {\rm deg} D_2 \geq l(D_1) \]
since otherwise we could find in $L(D_1)$ the function having zeros
at $Q_1,...,Q_l$ and $l(D)\neq 0$.

Moreover, by Riemann-Roch inequality we have
\[ l(D_1)\geq {\rm deg} D_1 -g +1 = {\rm deg} D +{\rm deg} D_2-g+1 \]
Combining this with the previous relation we have
\[ {\rm deg} D \leq g-1 \]
In full analogy using relation $l(C-D)=0$ we obtain
\[ {\rm deg} (C-D) \leq g-1 \]
or, using Corollary 4,
\[ {\rm deg} D\geq g-1 \]
So $\deg D =g-1$ and the Riemann-Roch theorem is completely proved.
$\Box$

\subsection{Riemann-Hurvitz formula}

Here we shall prove another useful fact - the {\it Riemann-Hurvitz
formula} that allows to calculate  the genus of algebraic curves.
For this purpose we shall use Corollary 4 which shows that degree
of an arbitrary divisor from canonical class is a purely topological
thing: it
depends only on the genus of algebraic curve.

\begin{theorem}[Riemann-Hurvitz]
Let an algebraic curve $\hat{\cal L}$ of genus $\hat{g}$ be an $N$-sheeted
covering with the branch points
$P_1,...,P_n \in {\cal L}$ of degree $k_1,...,k_n$ respectively
 of the curve ${\cal L}$ of genus $g$y.
Then
\begin{equation}
2\hat{g}-2 = N(2g-2)+\sum_{j=1}^{n} k_j
\label{RH1}
\end{equation}

(The order $k_j$ of the branch point $P_j$ is defined as the number of the
copies of the curve ${\cal L}$ which coalesce at this point minus 1)
\end{theorem}

{\it Proof.} The main idea of the proof is the following: starting with some
holomorphic differential on ${\cal L}$ one can construct explicitly some
holomorphic differential on $\hat{\cal L}$ and  calculate the degree of its
divisor.

So take on ${\cal L}$ some holomorphic differential $dU(P)$
(existence of $dU$ implies that $g\geq 1$; case $g=0$ may be considered
analogously taking an arbitrary meromorphic differential on ${\cal L}=
\bar{\bf C}$). The number of zeros of $dU$ on ${\cal L}$ is equal to $2g-2$.
Now let's define the differential $d\hat{U}(P),\;\;P
\in\hat{\cal L}$ by the following formula:
\[ d\hat{U}(P)=dU(\pi(P)) \]
where $\pi :\hat{\cal L}\rightarrow {\cal L}$ is the natural projection. So
we simply choose $d\hat{U}(P)$ taking the same value on all "sheets"
(i.e. copies of ${\cal L}$) of $\hat{\cal L}$. Obviously $d\hat{U}(P)$
has on $\hat{\cal L}$ $N(2g-2)$ zeros corresponding to the zeros of
$dU(P)$ on ${\cal L}$. Besides that, it has some additional zeros due to the
different local parameters on ${\cal L}$ and $\hat{\cal L}$ in the
neighbourhoods of the branch points $P_1,...,P_n$.

Namely, if we define  the local parameter  on ${\cal L}$ in the neighbourhood
of the point $P_j$ by 
$\tau$, then
the local parameter at the same point on $\hat{\cal L}$ is equal to
\[ \hat{\tau}=\sqrt[k_j+1]{\tau} \]
($k_j+1$ is the number of copies of ${\cal L}$ glued at $P_j$) or,
equivalently,
\[ \tau = \hat{\tau}^{k_j+1} \]
So if $dU(P)$ has in the neighbourhood of $P_j$ on ${\cal L}$ the local
representation
\[ dU(\tau)=f(\tau)d\tau \]
then for the differential $d\hat{U}(P)$ on $\hat{\cal L}$ we have
\[ d\hat{U}(\hat{\tau})=dU(\tau)=f(\tau)d\tau=f(\hat{\tau}^{k_j+1})
(k_j+1) \hat{\tau}^{k_j}d\hat{\tau} \]
i.e. $P_j$ is the zero of $d\hat{U}$ of degree $k_j$.
As a result, the number of zeros of differential $d\hat{U}$ on
$\hat{\cal L}$ is equal to
\[ N(2g-2)+\sum_{j=1}^{n} k_j =2\hat{g}-2 \]
where $\hat{g}$ is the genus of the curve $\hat{\cal L}$. $\Box$

or $g=0$ we get the following

\begin{corollary}
If curve ${\cal L}$ is $N$-sheeted covering of the Riemann sphere having $n$ 
branch
points of the order $k_1,...,k_n$
then the genus of ${\cal L}$ is equal to
\begin{equation}
g=\sum_{j=1}^{n} \frac{k_j}{2}-N+1
\label{RH3}
\end{equation}
\end{corollary}

For example, we can calculate the genus of the hyperelliptic curve (\ref{he}):
in this case $N=2$, $k_j=1,\;j=1,...,2g+2$. Substituting this into
(\ref{RH3}) we see that the genus is equal to $g$.

\section{Theta-functions on Riemann surfaces}

Here we present some facts about the one- and multi-dimensional
theta functions on Riemann surfaces that provide the universal tool
to explicitly construct the meromorphic functions and the functions with
exponential essential singularities.

\subsection{Definition and simplest properties}

\begin{definition}
Let $B$ be  symmetric $g\times g$ matrix with positively-defined
imaginary part. Then the function
\begin{equation}
\Theta(z|B)= \sum_{m\in {\bf Z^g}} \exp\{\pi i\langle Bm,m\rangle +2\pi i
\langle z,m\rangle\}
\label{theta}
\end{equation}
where $z\in{\bf C}^g$ and $\langle .,.\rangle $ is the
 ordinary scalar product, 
is called
the $g$-dimensional theta-function.
\end{definition}

The convergence of (\ref{theta}) for every $z\in {\bf C}^g$ immediately
follows from the condition that the matrix $Im B$ is positively defined.
The convergence is absolute and uniform on  every compact
domain in ${\bf C}^g$ and, therefore, $\Theta(z|B)$ considered
as a function of $z$ is holomorphic everywhere on ${\bf C}^g$.

If matrix $B$ is fixed, we shall often use the brief notation
\[ \Theta(z)\equiv \Theta(z|B) \]

Function (\ref{theta}) possesses important periodicity properties.
To formulate them denote by $e_1,...,e_g$ the standard basis in ${\bf C}^g$:
\[ (e_j)_k=\delta_{jk} \]
and by $f_1,...,f_g$ the columns of matrix $B$:
\[ f_j=Be_j\;\;,\;\;j=1,...,g \]

\begin{statement}
Theta-function (\ref{theta}) satisfies the following relations:

\begin{equation}
\Theta(z+e_j)=\Theta(z)
\label{pe}
\end{equation}

\begin{equation}
\Theta(z+f_j)=e^{-\pi i B_{jj}-2\pi i z_j} \Theta(z)
\label{pf}
\end{equation}

\end{statement}

{\it Proof.} The relation (\ref{pe}) is obvious. Let's prove (\ref{pf}):

\[ \Theta(z+f_k)=\sum_{m\in {\bf Z^g}} \exp\{\pi i\la Bm,m\ra 
+2\pi i\la m,z+f_k\ra \}\]
\[=\sum_{n\in {\bf Z^g}\;(m=n-e_k)} \exp\{\pi i\la B(n-e_k),n-e_k\ra +
2\pi i \la n-e_k,z+f_k\ra \}\]

\[ =\sum_{n\in{\bf Z^g}} \exp\{\pi i \la Bn,n\ra -2\pi i\la Be_k,n\ra 
+\pi i\la Be_k,e_k\ra \]

\[ + 2pi i \la n,z\ra +2\pi i\la n,f_k\ra -2\pi i \la e_k,z\ra 
-2\pi i\la e_k,f_k\ra \} \]

\[=\exp(-\pi i\la Be_k,e_k\ra -2\pi i\la e_k,z\ra )\Theta(z)=
 \exp(-\pi i B_{kk}-2\pi i z_k)\Theta (z) \;\;\;\Box \]

So the vectors $e_k$ are the periods of function $\Theta(z)$; the 
vectors $f_k$ are
called  the quasi-periods.

Now we immediately see that for arbitrary $N,M\in{\bf Z}^g$
\[ \Theta(z+N+BM)=\exp\{-\pi i\la BM,M\ra -2\pi i\la M,z\ra \} \Theta(z) \]
In fact the periodicity properties (\ref{pe}),(\ref{pf}) almost completely 
define $\Theta(z)$ up to some non-essential transformations {\cite{Mum}}.

The theta-function (\ref{theta}) admits 
natural generalization to the 
the theta-function with characteristics:
\begin{equation}
\Theta[\alpha,\beta](z|B)=\sum_{m\in{\bf Z^g}}
\exp\{\pi i\la B(m+\alpha\ra ,(m+\alpha)\ra +2\pi i\la z+\beta,m+\alpha\ra\}
\label{thetac}
\end{equation}
or, equivalently,
\[ \Theta[\alpha,\beta](z|B)=\exp\{\pi i\la B\alpha,\alpha\ra +
2\pi i \la z+\beta,\alpha\ra \}\Theta(z+\beta+B\alpha) \]
where $\alpha,\beta\in {\bf R}^g$.

The periodicity property of the theta-functions
with characteristics is the following:
\[ \Theta[\alpha,\beta](z+N+BM)= \]
\begin{equation}
=\exp\{-\pi i\la BM,M\ra -2\pi i\la z,M\ra +
2\pi i[\la \alpha,N\ra -\la\beta,M\ra ]\}\Theta[\alpha,\beta](z)
\label{trans}
\end{equation}
If vectors $\alpha$ and $\beta$ consist of $0$ and $\frac{1}{2}$ then the
set $[\alpha,\beta]$ is called the  half-period. Half-period $[\alpha,\beta]$
is called {\it even} if $\;4\la \alpha,\beta\ra \equiv 0(\bmod\;2)$ 
and {\it odd}
in the opposite case. For related theta-function we have the
following
\begin{statement}
Function $\Theta[\alpha,\beta](z)$ with half-integer characteristics is
even if half-period $[\alpha,\beta]$ is even and odd if half-period
$[\alpha,\beta]$ is odd.

\end{statement}

{\it Proof.} Substituting in (\ref{thetac}) $-z$ instead of $z$ and changing
the summation variable as $m\rightarrow -m-2\alpha$, we see that 
(\ref{thetac}) multiplies by the factor $\exp\{4\pi i\la\alpha,\beta\ra\}$.
 $\Box$

As the  simple corollary we see that 
\[ \Theta(z)=\Theta(-z) \]

Before to demonstrate how the meromorphic functions on the Riemann
surfaces of an arbitrary genus can be constructed in terms of the
multi-dimensional theta-functions let's consider the simplest case
of genus 1.

\subsection{Meromorphic functions on
elliptic curves in terms of theta-functions}

Four well-known elliptic Jacobi theta-functions have all possible sets
of half-integer characteristics $[\alpha,\beta]$:
\[ i\Theta_1(z)\equiv\Theta[\frac{1}{2},\frac{1}{2}](z) \]
\[ \Theta_2(z)\equiv \Theta[\frac{1}{2},0](z) \]
\[ \Theta_3(z)\equiv \Theta[0,0](z)\equiv \Theta(z) \]
\[ \Theta_4(z)\equiv \Theta[0,\frac{1}{2}] \]

All functions  $\Theta_i$ are holomorphic on ${\bf C}$ (of course, not on
$\bar{\bf C}$). Numbers $1$ and $B$ define the so-called parallelogram of
the periods.
Using the transformation property (\ref{trans}), it is easy to see that the
set of zeros
of  every function $\Theta_i$ is invariant with respect to the shift along  
every period.
Thus it is sufficient to find the zeros of $\Theta_i$ in the
parallelogram of the periods $\Omega$ (Fig. 6).
\vskip6.0cm
\begin{center}
{\bf Figure 6}
\end{center}
\vskip0.5cm
>From statement 1 it follows that function $\Theta_1(z)$ is odd
and other $\Theta_i$ are even. So $\Theta_1(0)=0$; the zeros of other
$\Theta_i$ may be easily found from (\ref{trans}). In particular, we see
that
\[ \Theta(\frac{B}{2}+\frac{1}{2})=0 \]
It is easy to verify that this zero of $\Theta(z)$ in $\Omega$ is unique.
Namely, consider the following integral along 
boundary $\partial \Omega$:
\[ \frac{1}{2\pi i}\oint_{\partial\Omega}\frac{d\Theta(z)}{\Theta(z)}=
\frac{1}{2\pi i}\oint_{\partial \Omega}d\log\Theta(z)= \]
\[=\frac{1}{2\pi i}\int_{0}^{1}[d\log\Theta(z)-d\log\Theta(z+B)]+
\frac{1}{2\pi i}\int_{0}^{B}[d\log \Theta(z+1)-d\log\Theta(z)]= \]
\[ =\frac{1}{2\pi i}\int_{0}^{1}\{d\log\Theta(z)-
d\log[e^{-\pi iB-2\pi iz}\Theta(z)]\}=
\frac{1}{2\pi i}\int_{0}^{1}2\pi i dz=1 \]

We have proved the following 
\begin{statement}
The elliptic function $\Theta(z|B)$ has in the parallelogram  $\Omega$ with
sides $1$ and $B$ only one zero at the point $z=\frac{1}{2}+\frac{B}{2}$.
\end{statement}

As we shall see below, this property together with the transformation
properties allows to express the
meromorphic functions on torus in terms of $\Theta(z|B)$.

We shall skip here numerous remarkable properties of the Jacobi
theta-functions such as summation theorems, Jacobi formula for the
derivative and so on (see {\cite{Chandra,Mum,Watson}}).
Notice only that
it obeys the
heat conductivity equation as the function of $z$ and $B$:

\begin{equation}
4\pi i\frac{\partial\Theta (z|B)}{\partial B}=
\frac{\partial^2\Theta (z|B)}{\partial z^2}
\label{hc}
\end{equation}
which may be easily verified by direct differentiation.

Now we are going to show how to construct functions on algebraic
curve of genus 1 in terms of elliptic function $\Theta(z|B)$.

First we  construct meromorphic functions on the abstract
Riemann surface - the 
torus $T={\bf C}/\{1,B\}$ ($T$ is in fact the fundamental
parallelogram
$\Omega$ with identified opposite sides); later we shall clarify
 the relationship between torus $T$ and an arbitrary elliptic
curve.

Following {\cite{Mum}} we can propose several ways to express meromorphic
functions on $T$ in terms of $\Theta(z|B)$:

{\bf 1.} Let's construct function $f(z)$ on $T$ having zeros
at the points $a_1,...,a_n$ and poles at the points
 $b_1,...,b_n$. Due to the Abel theorem
we can not, in contrast to the Riemann sphere, take arbitrary
values of $a_j$ and $b_j$: we get  one additional restriction given by
(\ref{At}). The unique holomorphic differential $dU(z)$ may be written on
$T$ in the very simple form:
\[ dU(z)=dz ;\]
so restriction (\ref{At}) gives  the following condition of  existence
of function $f(z)$:
\begin{equation}
a_1+...+a_n=b_1+...+b_n
\label{pz}
\end{equation}
We again see that it is impossible to construct non-trivial function $f(z)$
having one pole and one zero: in this case $a_1=b_1$ and $f(z)=const$.

For $n\geq2$ the function $f(z)$ may be expressed in terms of $\Theta(z|B)$ as
follows:
\begin{equation}
f(z)=C\prod_{j=1}^{n}\frac{\Theta(z-a_j-\frac{B+1}{2})}
{\Theta(z-b_j-\frac{B+1}{2})}
\label{m1}
\end{equation}
To verify that $f(z)$ is indeed singlevalued on $T$
we have to
check that $f(z+1)=f(z)$ and $f(z+B)=f(z)$. The first relation
is trivial; using the periodicity property (\ref{pf}) of the theta-function,
it is easy to see that the second of them reduces to (\ref{pz}).

Formula (\ref{m1}) may be considered as the straightforward generalization
of the following representation for the meromorphic (rational) function
on $\bar{\bf C}$ (${\bf CP}^1$):
\begin{equation}
f(z)=\prod_{j=1}^{n} \frac{z-a_j}{z-b_j}\;;
\label{fr}
\end{equation}
the most important difference is that in (\ref{fr}) the positions of the
poles and zeros
are arbitrary.

Another way to construct meromorphic functions on $T$ in terms of the
theta-functions is the following:

{\bf 2.} Consider the function $\log \Theta(z)$. It is obviously equal
to the sum of some periodic function with  periods $(1,B)$ and some linear
function. So function $\frac{d^2}{dz^2} \log \Theta(z)$ is meromorphic
and periodic, and, therefore, is a meromorphic function on $T$. Besides that,
it has the double pole at $z=\frac{1+B}{2}$.
This is nothing but the
 Weierstrass ${\cal P}$-function up to some constant:
\[ {\cal P}(z)=-\frac{d^2}{dz^2} \log\Theta(z) + const \]
where ${\it const}$ is chosen to kill the term of zero degree in Laurant
series of ${\cal P}(z)$ at $z=0$.

{\bf 3.} Finally, by slight modification of the previous method we can
construct the an arbitrary meromorphic function on $T$ in the following
form:
\begin{equation}
f(z)=\sum_{j} \lambda_j\frac{d}{dz}\log\Theta(z-b_j) + const
\label{m3}
\end{equation}
For the function (\ref{m3}) be singlevalued on $T$, we have to impose the
condition
\[ \sum_{j} \lambda_j =0 \]
that provides the condition of Abel theorem in this case. Notice that in
(\ref{m3}) we have no apparent information about the zeros of $f(z)$;
instead we know the residues $\lambda_j$ at the points $b_j$.

Representation (\ref{m3}) is the analog of the representation of meromorphic
function on ${\bf CP^1}$ as the sum of the simple fractions:
\[ f(z)=const + \sum_{j}\frac{\lambda_j}{z-b_j} \]
Now to be able to construct meromorphic functions on an arbitrary
elliptic curve one should be able to establish an isomorphism
between the algebraic curve and the fundamental parallelogram with some $B$.

This isomorphism is defined by the Abel map
\begin{equation}
U(P)=\int_{P_0}^{P} dU
\label{Am1}
\end{equation}
where $dU$ is unique normalized holomorphic differential on ${\cal L}$;
the first period of the Jacobi torus is equal to
\[ \oint_{a}dU=1 \]
and the second period $B$ should coincide
with the second period of the torus $T$. It may be easily verified that in
the case $g=1$ the correspondence between ${\cal L}$ and $J({\cal L})$ is
an the isomorphism (for $g>1$ this is, of course, not true, because
$\dim J({\cal L})=g$). The direct correspondence ${\cal L}\rightarrow
J({\cal L})$ is
given by the Abel map (\ref{Am1}); the inverse is defined by the Weierstrass
${\cal P}$-function {\cite{Chandra,dub}}.

Now being able to construct meromorphic functions on $J({\cal L})$ by the
methods {\bf 1.-3.} we can easily construct meromorphic functions
on ${\cal L}$.

For example, according to the method {\bf 1.}, the function $f(P)$ on ${\cal L}$
having poles at points $P_1,...,P_n$ and zeros at $Q_1,...,Q_n$ related by
the Abel theorem
\begin{equation}
\sum_{j=1}^{n} U(P_j)=\sum_{j=1}^{n} U(Q_j)
\label{Atheo}
\end{equation}
may be defined by the following expression:
\[ f(P)=C\prod_{j=1}^{n}\frac{\Theta(U(P)-U(Q_j)-\frac{B+1}{2})}
{\Theta(U(P)-U(P_j)-\frac{B+1}{2})} \]

Certainly we can not usually explicitly resolve
 condition (\ref{Atheo}). 
However, it is not always necessary to know {\it all} the poles
and zeros of $f(P)$ on ${\cal L}$.

For example, function having on ${\cal L}$ 2 poles at $P_{1,2}$, one zero
at $Q_1$ and one zero somewhere also (we can not find it explicitly) may be 
expressed as follows:
\[ f(P)=\frac{\Theta(U(P)-U(Q_1)-\frac{B+1}{2})
\Theta(U(P)+U(Q_1)-U(P_1)-U(P_2)-\frac{B+1}{2})}
{\Theta(U(P)-U(P_1)-\frac{B+1}{2})\Theta(U(P)-U(P_2)-\frac{B+1}{2})} \]

The position of the second zero is given by the Abel theorem (\ref{Atheo}).

\subsection{Meromorphic functions on algebraic curves of arbitrary genus
in terms of theta-functions}

To extend the results of the previous paragraph to
 the curves of an arbitrary genus
 we have to prove some facts
about the zeros of the multidimensional theta-functions.

Consider the function
\begin{equation}
F(P)=\Theta(U(P)-d|B)
\label{FP}
\end{equation}
where $U(P)$ is the Abel map on ${\cal L}$; $B$ is the matrix of $b$-periods
and $d\in {\bf C}^g$ is some fixed vector. Of course, $F(P)$ is
non-singlevalued on ${\cal L}$; so consider it on the fundamental polygon
$\tilde{\cal L}$ where it is single-valued and holomorphic.
\begin{lemma}
If $F(P)$ is not identically zero 
 then it has on $\tilde{\cal L}$
$g$ zeros (taking into account their order).
\end{lemma}

{\it Proof.} We have to calculate the integral
\[ (\ast)=\frac{1}{2\pi i}\oint_{\partial \tilde{\cal L}} d\log F(P) \]

Let's use the representation (\ref{bfp}) for $\partial \tilde{\cal L}$.
Denote the value of $F(P)$ on $a_j^-$ and $b_j^-$ by $F^-$ and on
$a_j^+$ and $b_j^+$ - by $F^+$; the same for the Abel map $U(P)$.
Then
\begin{equation}
(\ast)=\frac{1}{2\pi i}\sum_{k=1}^{g} \left(\oint_{a_k}+\oint_{b_k}\right)
[d\log F^+ -d\log F^-]
\label{ast}
\end{equation}

Notice that if $P\in a_k$ then
\begin{equation}
U_j^-(P)=U_j^+(P)+B_{jk}
\label{U1}
\end{equation}
and if $P\in b_k$ then

\begin{equation}
U_j^+(P)=U_j^-(P)+\delta_{jk}
\label{U2}
\end{equation}
Using the periodicity properties of theta-function (\ref{pe}),
(\ref{pf}), we see that on $a_k$
\begin{equation}
d\log F^-(P)=d\log F^+(P)- 2\pi i dU_k(P)
\label{F1}
\end{equation}
and on $b_k$
\begin{equation}
d\log F^-(P)=d\log F^+(P)\;;
\label{F2}
\end{equation}
therefore integral (\ref{ast}) may be written as follows:
\[ (\ast)=\frac{1}{2\pi i}\sum_{k}\int_{a_k}2\pi i dU_k(P)=g \]
$\Box$

The information about the positions of the zeros of $F(P)$ on
$\tilde{\cal L}$ is given by the following
\begin{lemma}
Assume that $F(P)$ is not identically zero
on $\tilde{\cal L}$ and denote by $P_1,...,P_g$ the positions of its zeros.
Then on the Jacobi manifold $J({\cal L})$ we have
\[ \sum_{k=1}^{g} U(P_k)\equiv d-K\;\;(\bmod I,B) \]
where $K$ is the vector of the Riemann constants:
\begin{equation}
K_j=\frac{1+B_{jj}}{2}-\sum_{k\neq j}\oint_{a_k}U_j(P)dU_k(P)
\label{RK}
\end{equation}
\[ j=1,...,g \]
\end{lemma}

{\it Proof.} Consider the integral
\[ I_j=\oint_{\partial \tilde{\cal L}} U_j(P)d\log F(P),\;\;\;j=1,...,g \]

>From the residue theorem we have
\[ I_j=\sum_{k=1}^{g} U_j(P_k) \]

Now calculate $I_j$ in analogy to the previous lemma using
relations (\ref{U1}),(\ref{U2}),(\ref{F1}),(\ref{F2}):
\[ I_j=\frac{1}{2\pi i}\sum_{k=1}^{g} \left(\oint_{a_k}+\oint_{b_k}\right)
[U_j^+d\log F^+ -U_j^- d\log F^- ] \]
\[ =\frac{1}{2\pi i}\sum_{k=1}^{g} \oint_{a_k}[U_j^+ d\log F^+ -
(U_j^+ +B_{jk})(d\log F^+ -2\pi i dU_k)] \]
\[ +\frac{1}{2\pi i}\sum_{k=1}^{g}\oint_{b_k}[U_j^+ d\log F^+ -
(U_j^+ -\delta_{jk})d\log F^+] \]
\[ = \sum_{k=1}^{g}\left(\oint_{a_k}U_j^+ dU_k -\frac{B_{jk}}{2\pi i}\oint_{a_k}
d\log F^+ +B_{jk}\right) +\frac{1}{2\pi i}\oint_{b_j}d\log F^+ \]

Consider these integrals separately:
\[ \oint_{a_k} d\log F^+ =2\pi i n_k\;,\;\;\;n_k\in {\bf Z} \]
since function $F^+$ takes the same values at the different ends of $a_k$.

Let $Q_j$ and $\tilde{Q}_j$ be the beginning and the end of $b_j$. Then
\[ \oint_{b_j} d\log F^+ =\log F^+(\tilde{Q}_j)-\log F^+(Q_j) +2\pi i m_j \]
\[ =\log\Theta(U(Q_j)+f_j-d)-\log\Theta(U(Q_j)-d) +2\pi i m_j  \]
\[ =-\pi i B_{jj}+2\pi i d_j -2\pi i U_j(Q_j) +2\pi i m_j\;,\;\;\;
m_j\in{\bf Z}, \]
vector $f_j$ is the same as in (\ref{pf}).

As a result we have module the lattice periods
\[ I_j\equiv d_j-\frac{B_{jj}}{2}-U_j(Q_j)+\sum_{k=1}^{g}\oint_{a_k}
U_j(P)dU_k(P) \]
Denoting the beginning of contour $a_j$ by $R_j$ (its end coincides with
$Q_j$), we have
\[ \oint_{a_j}U_j(P)dU_j(P)-U_j(Q_j)=\frac{1}{2}[U_j^2(Q_j)-U_j^2(R_j)]-
U_j(Q_j) \]
\[= \frac{1}{2}[(U_j(R_j)+1)^2-U_j^2(R_j)]-U_j(R_j)-1=-\frac{1}{2} .\]

Finally,
\[ I_j=\sum_{k=1}^{g} U_j(P_k)=d_j-\frac{B_{jj}+1}{2} +\sum_{k=1,\;k\neq j}
^{g}\oint_{a_k}U_j(P)dU_k(P) \]
$\Box$

Notice that the vector $K$ is in fact not the "constant" - it depends on the
choice of the initial point of the Abel map $P_0$; this dependence
disappears only for $g=1$; in this case we have
\[ K=\frac{B+1}{2} \]

It appears possible to prove {\cite{Grif}} the following simple link
between vector $K$ and the divisors from the canonical class $C$ (i.e. the
divisors of meromorphic differentials):
\[ 2K\equiv-U(C) \]

Expression (\ref{RK}) may be considerably simplified {\cite{Fay}} in the
case of the hyperelliptic curves if the point $P_0$ is 
chosen in some special way:
\begin{lemma}
Let ${\cal L}$ be  hyperelliptic curve of genus $g$ defined by the equation
(\ref{he}) and $\sigma :{\cal L}\rightarrow {\cal L}$ be the involution on
${\cal L}$ interchanging the sheets. Let also the canonical basis of
cycles $(a_j,b_j)$ be chosen in such a way that $\sigma(a_j)=-a_j$ and
$\sigma(b_j)=-b_j$ (Fig.7). Then choosing  the point $P_0$ to coincide with 
$E_1$ we can express the
vector of the Riemann constants as follows:
\begin{equation}
K_j=\sum_{k=1}^{g} B_{jk} +\frac{j}{2}
\label{RKh}
\end{equation}
\end{lemma}

{\it Proof.}
Existence of the involution $\sigma$ allows to rewrite expression
(\ref{RK}) as follows:
\[ K_j=\frac{B_{jj}+1}{2}+\sum_{k\neq j} \oint_{a_k}\left[
\int_{P_0}^{P}dU_j\right]dU_k=
 \frac{B_{jj}+1}{2}+\sum_{k\neq j}\oint_{a_k}\left[\int_{E_1}^{E_{2k+1}}dU_j+
\int_{E_{2k+1}}^{P}dU_j\right]dU_k(P) \]
\[ =\frac{B_{jj}+1}{2}+\sum_{k\neq j}\int_{E_1}^{E_{2k+1}}dU_j\oint_{a_k}
dU_k  \]
\[+\sum_{k\neq j}\int_{E_{2k+1}}^{E_{2k+2}}\left(\left[\int_{E_{2k+1}}^{P}
dU_j\right]dU_k(P)-\left[\int_{E_{2k+1}}^{\sigma P}\sigma^{\ast} dU_j\right]
dU_k(\sigma P) \right) \]
\[ \]
>From the relation $dU_j(\sigma P)=-dU_j(P)$ that immediately follows from the
behaviour of basic cycles under the involution $\sigma$ we see that the
second sum is equal to zero; the first sum gives (\ref{RKh})  module  the
linear combination $M+BN$. $\Box$

So we know the positions of the zeros of the theta-function on ${\cal L}$
if it is not identically zero. At this point we refer to the
following theorem (see {\cite{Mum,Hur}}):
\begin{theorem}
Function $\Theta(U(P)-d)$  identically vanishes on ${\cal L}$ iff
vector $d$ may be represented in the form
\[ d\equiv U(Q_1)+...+U(Q_g)+K \]
where $Q_1+...+Q_g$ is some special divisor.
\end{theorem}

As a result we get from Lemma 4 and Theorem 10
the following

\begin{theorem}
If $D=P_1+...+P_g$ is a non-special divisor on algebraic curve ${\cal L}$
of genus
$g$ then the function $F(P)=\Theta(U(P)-U(D)-K)$ has on $\tilde{\cal L}$
$g$ zeros at points $P_1,...P_g$.
\end{theorem}

Notice also the following important role of the zeros of theta-function.
In the case $g=1$ we have the one-to-one correspondence between the
curve ${\cal L}$ and its Jacobi torus $J({\cal L})$ given in different
directions by the elliptic
integral $U(P)$ and by Weierstrass ${\cal P}$-function.
For $g>1$ ${\cal L}$ and $J({\cal L})$ have different
complex dimensions and it is natural to consider a map between
${\cal L}^g$ and $J({\cal L})$. In analogy to the elliptic case we can define
the Abel map
\begin{equation}
U:{\cal L}^g \rightarrow J({\cal L})
\label{Am}
\end{equation}
that assignes to every positive divisor of degree $g$ some point
of $J({\cal L})$. The reasonable question here is how to construct
(if possible) the inverse map i.e. to solve the equation
\[ U(D)\equiv \zeta \]
where $\zeta\in {\bf C}^g$ is some point on $J({\cal L})$; we want to
find the divisor $D=P_1+...+P_g$ in terms of $\zeta$. This is the
formulation of the Abel inversion problem.

Theorems 10 and 11 provide the answer to this question: if
$\Theta(U(P)-\zeta-K)$ does not vanish identically on ${\cal L}$
then its zeros coincide with the 
points $P_1,...,P_g$ which solve the Abel inversion problem. We shall not
consider in details the particular case $\Theta(U(P)-\zeta-K)\equiv 0$;
in this case according to Theorem 11
$\zeta=U(\tilde{D})$ where divisor $\tilde{D}$ is special.
It is easy to prove {\cite{dub,Mum}} that all these points $\zeta$ may
be represented as
\[\zeta=U(Q_1+...+Q_{g-1}) \]
where $Q_i,\,i=1,...,g-1$ are arbitrary points of ${\cal L}$.
Taking into account that the special divisors constitute the submanifold in
${\cal L}^g$ of complex dimension $g-1$, it is possible to claim that the
Jacobi inversion problem may be solved for any $\zeta\in J({\cal L})$
and Abel map gives the one-to-one correspondence between
 ${\cal L}^g$ and
$J({\cal L})$ everywhere except the special divisors on ${\cal L}$.

Now we are in position to achieve our last goal  - to demonstrate how
we can construct the meromorphic functions on ${\cal L}$ in terms of the
theta-functions. Here, in analogy to the elliptic case, there are many
possibilities.
For illustration let's  construct function $f$
having poles in prescribed $g+1$ points $P_1,...,P_{g+1}$ (in general
case, when divisor $P_1+...+P_{g+1}$ is non-special, this function is
unique up to a linear transformation).

\begin{lemma}
Let $P_1+...+P_{g+1}$ be a non-special divisor on the curve ${\cal L}$
of genus $g$. Then the function
\begin{equation}
f(P)=\frac{\Theta(U(P)-\sum_{j=1}^{g+1}U(P_j)+U(Q)-K)}
{\Theta(U(P)-\sum_{j=1}^{g}U(P_j)-K)}\exp(W_{QP_{g+1}})\;\;\;,
\label{mf}
\end{equation}
(where $Q\in {\cal L}$ and $W_{QP_j}$ is the  normalized (all $a$-periods are
zero) differential of the 3rd kind) is meromorphic on ${\cal L}$, has poles
at $P_1,...,P_{g+1}$ and zero at $P=Q$ (other zeros are the zeros of the
theta-function in the numerator).
\end{lemma}

{\it Proof.} The fact that the expression (\ref{mf}) has poles at
$P_1,...P_g$ is obvious: $g$ poles come from the denominator
(here we use the assumption that the divisor $D$ is non-special) and pole
$P_{g+1}$ comes from the exponential factor; this factor
gives also the zero at $Q$. The only fact one should check is that
function $f(P)$ is singlevalued on ${\cal L}$ i.e. that it is invariant
with respect to the
tracing of the point $P$ around any of the basic cycles.
The invariance of $f$ with respect to the tracing around $a$-cycles is obvious
due to the normalization of the differential
$dW_{QP_{g+1}}$ and the periodicity of the theta-function
(\ref{pe}). The invariance with respect to the tracing around 
$b$-cycles follows
from the expression (\ref{b3}) for the vector of $b$-periods of the 
integral of the
third kind and the transformation property of the theta-function (\ref{pf}).
$\Box$

Analogously we can construct  meromorphic functions with more poles;
it also appears possible to express in terms of the theta-functions 
arbitrary differentials of the third and second kind
through the so-called prime form - the special differential of
the order $(-\frac{1}{2},-\frac{1}{2})$ on ${\cal L}\times {\cal L}$
(see {\cite{Mum,bim}})

Another opportunity is to insert in the exponent of (\ref{mf}) the normalized
integral of the 2nd kind instead of $dW_{QP_{g+1}}$ (adding simultaneously
its $b$-period vector divided by $2\pi i$ in the argument of the
theta-function in the numerator). Then we get the function having on ${\cal L}$
$g$ poles and an essential singularity of the exponential kind. The
functions of this sort with one or more essential singularities
(so-called Baker-Akhiezer functions) arise in
applications to  KdV-like  soliton equations.
The construction of the functions of this kind in terms of the theta-functions
is considered in details in {\cite{bim}}.
The pure meromorphic functions arise in the 
applications, for example,  to the
Ernst  equations \cite{KOR}.

\end{document}